\documentclass[reprint,longbibliography,pre,amsmath,amssymb,amsfont,url, aps]{revtex4-2}
\usepackage{bold-extra}
\usepackage{amsthm}
\usepackage{dutchcal}

\usepackage[colorlinks, allcolors=blue]{hyperref}
\usepackage{graphicx} 
\usepackage{csquotes}
\usepackage{mathtools}

\usepackage{enumitem}
\setlist[enumerate]{parsep=3pt}

\newcommand{\myquote}[1]{\textit{\textquote{#1}}}
\newcommand*\diff{\mathop{}\!\mathrm{d}}
\newcommand{\law}[2]{
	\vspace{0.25cm}
	\noindent
	\textbf{#1}: 
	\textit{#2}
	\vspace{0.2cm}
}
\begin{document}
	\author{D. Lairez}
	\email{didier.lairez@polytechnique.edu}
	\affiliation{Laboratoire des solides irradi\'es, \'Ecole polytechnique, CEA, CNRS, IPP,
		%Institut Polytechnique de Paris,
		91128 Palaiseau, France}
	\title{The principles of neopositivism and the laws of thermodynamics}
	\date{\today}
	
	\begin{abstract}
The second law of thermodynamics, which deals with irreversibility and makes the theory so special, is usually considered empirical. The definition of equilibrium as an attractor, on the other hand, requires a postulate. This article shows that both are actually already contained, even if hidden, in the fundamental principles of neopositivism, which are widely accepted in all fields of science.
In particular, from the definition of information as a truth that can only come from an observation but cannot be redundant, we obtain Clausius' inequality.
	\end{abstract}
	
	\maketitle
	
	\section*{Introduction}
	
Symmetries of an experiment in physics are the possible transformations that leave its result unchanged. They are those that are undetectable by simply observing the result of the experiment. 
A symmetry leaves invariant certain observed regularities of the world and the laws of physics that account for them\,\cite{Feynman_1963_symmetry}.

For instance, rotation by a given angle is a symmetry for all the laws of physics. But, rotation at a given angular velocity is not. 
Translation in space is a universal symmetry. 
Translation in time is also a universal symmetry, but time reversal is not a universal symmetry. 
And this last point is very confusing and annoying for physicists. What is the problem? It lies in the fact that all the fundamental laws of physics, such as those of electromagnetism and mechanics (classical, relativistic or quantum), are invariant under time reversal, with the exception of the second law of thermodynamics concerning irreversibility. 
This therefore represents a major challenge, as it is likely that the unification of fundamental interactions (that is, the unification of relativistic and quantum mechanics) will not be sufficient to arrive at the mythical theory of everything. The unification of physics will also require taking the laws of thermodynamics into account. Any new insight into these laws is therefore welcome. 

In physics, and more broadly in all scientific fields, statements of a theory belong to two categories\,: laws and principles\,\cite{Poincare_1905_Sci_Hypo}, also called empirical laws and theoretical laws\,\cite{Carnap_1966}, which are of a different nature. The former come from induction and are simply the generalization of particular observations. 
They express observed regularities of the world and are true until proven otherwise.
The latter concern the rest, that is to say everything that cannot be observed, such as the definition of concepts used to express empirical laws and the way in which these concepts articulate. They are actually conventions or postulates.
They are neither true nor false (because they cannot be faced to experiments), they are simply convenient or not with regard to the object of science, namely describing reality which itself remains to be defined by convention.
Principles in physics function fundamentally like axioms in mathematics, except that they are not arbitrary but have pragmatic obligations\,\cite{Poincare_1907}.

The first convention preceding any theory concerns the very object and purpose of science. Since science explicitly refers (by its etymology) to knowledge, this first convention also concerns what we mean by "knowledge". The aim of this article is to show that if we adopt the neopositivist\,\cite{Neurath1973} conventions on these points (in short\,: there is no synthetic \textit{a priori} knowledge other than that of logic; the sole object of science is reality understood as that which can be observed) the empirical second law about irreversibility can be deduced. In other words, the empirical second law about irreversibility is already contained in a set of fundamental principles which are common to all fields of physics\,: \myquote{the scientific conception of the world}\,\cite{Neurath1973}.

Our derivation places the concepts of knowledge, information and data in a fundamental position. A piece of information is a truth that cannot be redundant but is necessarily carried by data understood as a piece of observation. 
From this asymmetrical role of information and data  originates thermodynamic irreversibility.

The article is organized as follows\,: in the first section the fundamental principles of neopositivism are presented; the second section deals with the second law of thermodynamics (the first on energy conservation not being specific to it) and the definition of equilibrium; finally in the third section we will see how these laws of thermodynamics are already contained in the fundamental principles.

\section{Fundamental principles of neopositivism}\label{neo}

It is impossible to do science without conventions, also called postulates or principles, at least those that concern its object matter and what we mean by explaining and understanding. These epistemological conventions are often overlooked, which leads to useless debates simply because people are not talking about the same thing. This will be avoided by stating them from the outset. 
My personal observation of the scientific practice shows that these conventions are widely shared and should not offend many people. However, even if this were the case, these people should not deny that accepting these conventions implies certain things, which is precisely the subject of this article.
These conventions can be considered as axioms in mathematics; our purpose is not to discuss their validity (that would be a philosophical debate on the foundations of science; in my opinion, they are, but that is not the subject).
Here, our purpose is rather what they imply. For this reason, they are listed below as affirmative and undisputed statements (principles or definitions) that must be understood within the context of neopositivism.

These conventions are those of the Vienna Circle\,\cite{Neurath1973} and of its neighbourhood\,\cite{Wittgenstein1922, Carnap_1966, Mach_1911, Mach_1919, Poincare_1905_Sci_Hypo, Poincare_1907, Einstein_1934}, namely the neopositivism, or logical positivism, or logical empiricism. It is briefly presented below in a way that may seem unusual to readers who are already familiar with the original work of the Vienna Circle\,\cite{Neurath1973}. The reason is simply to adapt the presentation to our objective. We will talk in particular about information, a concept that was not part of the Vienna Circle's arsenal, but which we will need, and which will be treated consistently with neopositivist thinking. In addition, certain inconsistencies in the original work are corrected. These concern the role of conventions which are neither true nor false statements. The Circle explicitly acknowledges this role, but does not draw the necessary conclusion from it\,: logic must be three-valued. The consequence of this is crucial because it opens the door to the use of \textit{a priori} probabilities (this point will become clearer later).
\\

The only thing that science can talk about is scientific reality. In addition \myquote{whereof one cannot speak, thereof one must be silent} (Wittgenstein\,\cite{Wittgenstein1922}), so that there is no other reality in scientific statements. It is defined as\,:

\law{Definition (reality)} {Reality is what can be observed, with observation being understood as an interaction involving the observer.}

This reality is not made up of objects that could exist independently of us. Reality is inseparable from the observer. This has the consequence, in particular, of requiring us to define what we mean by \textquote{objectivity}. 
It is no longer the property of an object, but that of an observation which must be reproducible by anyone and which then becomes true by definition.

\law{Principle (objectivity)} {An objective observation can be reproduced by anyone and is true by definition.}

\noindent In what follows, we will only consider objective observations and omit the adjective. 

We expect science to help us understand reality, and understanding means connecting observations together. These connections can be of different kinds corresponding to different levels of understanding. It is agreed that

\law{Principle (understanding)}{The highest level of understanding is reached when all observations, past and future, can be deduced from a finite set of statements that forms a theory.}

The set of statements of a theory is intended to be used by logic (hence the name logical-positivism) and must therefore contain no contradictions. Also, we assume the principle of non-contradiction for a simple statement\,:

\law{Principle (non-contradiction)}{A statement cannot be both true and not true.}

Any statement is judged with respect to its truth.
A statement is true if it relates (narrates) an observation, a fact. In other words, a statement is true if it agrees with all observations. A statement is false if it disagrees with an observation. True and false statements are said to be \textit{a posteriori}. 
\textit{A priori} statements are those which are independent of any observation, such as conventions or definitions. They are neither true nor false. It follows that:\,

\law{Principle (truth 1/2)}{The only source of true statements (in short truths) is observations.}

\noindent Let us clarify this point. From a statement, others can be logically deduced, but \myquote{all [...] inference consists of nothing but a transition from statements to other statements that contain nothing that was not already in the former (tautological transformation).}\,\cite{Neurath1973}.
The statements thus produced cannot be considered a source of truth. Mathematics is tautological by nature. 

What is implicit in classical logic, whether it is bi-valued (two possible values: true or false) or three-valued (three possible values: true, false or unknown), is that the truth of a statement is a qualitative property (a quality) that this statement either possesses or does not possess. It is not a quantitative property (a gradual quantity).
That is, a true statement cannot be truer than another true statement. Furthermore, a true statement cannot become increasingly true even if it is repeated many times. For our purpose, this point deserves to be explicit\,:

\law{Principle (truth 2/2)}{The truth of a statement is a quality, not a gradual quantity.}

The statements of the theory are divided into two categories\,:
1)~\textit{A priori} statements, called principles, or postulates, or conventions. Neither true nor false, they are convenient. Their convenience is judged by whether or not they allow us to deduce observations from theory;
2)~True \textit{a posteriori} statements resulting from generalization of observed regularities.
They are called inductive or empirical laws, simply called laws in the following. 
Laws are true until proven otherwise by the observation of a counter-example. 

Truth and knowledge are linked. In philosophy, the definition of a knowledge is generally tripartite, it is a justified true belief\,\cite{Ichikawa2026}. 
With neopositivism, a truth can only come from an observation which is also its justification, \textquote{justified truth} is a pleonasm, there is no \textit{a priori} truth. A particular knowledge is just a true belief. That is, to know something is to be aware of the truth of that thing. 
By \textquote{knowledge of the observer} in general, we mean the sum total of truths (true statements) that he has in mind (in memory).

\law{Definition (knowledge)} {The knowledge of the observer is the sum total of truths he has in mind.}

\noindent Knowledge is therefore a state quantity of the observer. How it can be quantified is precisely the contribution of information theory, which we will discuss below. So, what is information? According to the Oxford English Dictionary, information is \myquote{the imparting of knowledge}. We will essentially adopt this common-sense definition, but with a reformulation better suited to what follows and which takes into account what has already been said.
While our knowledge is a set of truths stored in our mind, an information is a truth that is transmitted to it (via an observation/interaction) and increases our knowledge. Much like mechanical work or heat which are transmitted to a body but stored in the form of internal energy.

\law{Definition (information)}{An acquired piece of information (in short an information) is a truth transmitted to the observer via an observation/interaction.}

\noindent The truth of a statement is a quality that the statement either possesses or does not possess, but which cannot increase as the statement is duplicated, or as the observation that is its source is reproduced identically. Repeating a truth that is already known does not increase knowledge. So that,

\law{Principles (no redundancy)}{Redundant transmitted truths do not increase knowledge and counts as one single piece of information. An information is something new.}

\noindent This principle of non-redundancy, which is central in what follows, is sometime adopted in scientific literature (see e.g. \cite{Lutz_2015, Ciliberto_2018}), but not always. In particular, Landauer\,\cite{Landauer_1961} erases data-bits (that are transmitted truths that have been stored in mind), but he assimilates them to information-bits. From this confusion originates his famous \textquote{principle}. This point is widely discussed in a previous paper\,\cite{Lairez2025}. As for the aim of this article, let us remember that it is not about discussing the principle of non-redundancy, but simply about examining its implications.

In the framework of neopositivism, from the definition of reality it comes that there are no parallel universes. There is no Platonic intelligible and visible world. There is no spiritual and material world. There is no dualism. The universe is one. 
%Deducing things on a particular problem from a set of statements is essentially like finding the solution to an equation.
The deduction on a particular problem from the theory must also be unique (or univocal). But it is like finding the solution to an equation, the solutions are often multiple.
Randomness and probabilities are there precisely to address this problem. The solution (the deduction provided by the theory) is then a unique probability distribution.

\law{Principle (univocality)}{On a particular problem, deductions from the theory must be univocal in terms of probability distribution.}

Probabilities introduce randomness. 
Whether randomness is inherent to the nature of things or due to our ignorance leads to exactly the same observations and goes beyond the scope of science.
Traditionally, two types of probabilities are distinguished\,: \textit{a priori} and \textit{a posteriori}.
As indicated by their name, the first are completely detached from observation. They are actually conventions. Whereas, the second are relative statistical occurrences of observations. They are empirical laws. Both are mathematically treated in the same manner and both are valid.
But what is not permitted is 
either to say that 
the probabilities are \textit{a priori} distributed in a certain way, when they have been measured to be distributed in another way (this would be false);
or to say that 
the probabilities are \textit{a posteriori} without having been measured or without the possibility of doing so (this would be inconsistent).

An infinite set of possible observations is deduced from a finite set of statements and, in a certain sense, is contained within it (mathematics is tautological).
Theory is therefore a summary of reality. It is an economy of thought\,\cite{Mach_1911} and the most economical is the best. This is known as the Occam's razor\,\cite{Hahn1980}.
We are therefore faced with an optimization problem\,: \myquote{The basic concepts and laws which are not logically further reducible constitute the indispensable and not rationally deducible part of the theory. It can scarcely be denied that
	the supreme goal of all theory is to make the irreducible basic elements as simple and as few as possible without having to surrender the adequate representation of a single datum of experience} (Einstein\,\cite{Einstein_1934}).
The minimization procedure applies to the number of statements of the theory (conventions and laws), but it also applies to the laws themselves.
An inductive law is similar to inter- or extra\-polation of experimental measurements (observations). On the one hand, the law has the constraint to account for all observations, to account for everything that is true, to account for all knowledge.
But on the other hand, it must be as simple as possible. The best law must provide the \textquote{minimum service} within the imposed constraint of our knowledge. These two ideas about what is optimum, what is the \myquote{best}, what is the \myquote{supreme goal} (for the number of laws and for the laws themselves) are actually conventional value judgments. They can be expressed all in one under the form of an additional postulate\,:

%\law{Principle (least-talker)}{Theory must say the minimum about what we don't know, with the constraint of saying everything about what we know.}

\law{Principle (least-talker)}{Theory must minimize what is said (maximize the uncertainty) about what we do not know, with the constraint of saying everything about what we know.}

%\noindent Or, equivalently in terms of maximization\,:

%\law{Principle (highest-uncertainty)}{Theory must maximize the uncertainty about what we don't know, with the constraint of saying everything about what we know.}

\noindent Here is our final postulate, which sounds like an echo of the first\,: \myquote{Whereof one cannot speak, thereof one must be silent} (Wittgenstein\,\cite{Wittgenstein1922}). For this reason , the list is certainly not logically irreducible, but by being less concise it is considered clearer. 

\section{Statements of thermodynamics}\label{thermo}

By thermodynamics, we mean the initial phenomenological theory (Clausius\,\cite{Clausius_1879}), which deals only with the macroscopic observable behavior of a system and not with its internal functioning. 

Thermodynamics deals with the changes in form of energy. The first statement of the theory is therefore that of its conservation. This statement is actually a convention\,\cite{Poincare_1905_Sci_Hypo, Feynmann_Energy} and more precisely the definition of the concept itself. This principle is not specific to thermodynamics and is in fact common to all of physics; it will therefore not be addressed here and the reader is invited to consult a previous article\,\cite{Lairez2025}.

In thermodynamics, a system is described by state quantities and the state of a system is a way of being that is supposed to be maintained during a certain duration. Equilibrium is the stationary state for which this duration is not limited and is actually the only object matter of thermodynamics (\cite{Callen_1985} p.15). 

But there is a form of paradox in the intentions. How can we account for the transformations a system can undergo (the changes in form of energy) if we only deal with stationary states\,? The solution lies in the notion of quasi-static transformation and that of reversibility.
Another problem arises\,: a system out of equilibrium always tends towards equilibrium. Equilibrium is an attractor and this cannot be deduced from its definition as a stationary state.

Reversibility (thus irreversibility) and attractiveness of the equilibrium, are accounted for with the concept of entropy\,\cite{Clausius_1879}, which will be the entry point for the notion of information in the problem.
So let us start there.

\subsection{Entropy and reversibility}\label{SecLaw1}

In thermodynamics, a reversible cycle is a path that leads the system from an initial state through a sequence of other states and finally returns to the starting point, leaving everything (the system and its environment) exactly as it was before. Any process that can be part of such a cycle is said to be reversible. For example, a quasi-static process  sufficiently slow relative to any relaxation of the system is reversible. A quasi-static path is a succession of (quasi)equilibrium. It is under control and can be stopped at any time and then resumed or reversed to return to the starting point.
Note that, from the outset, we can clearly see the role that knowledge and information will play in thermodynamics\,: identifying whether everything is as before or being able to control a parameter depend on our knowledge. 

Clausius\,\cite{Clausius_1865} observed that for any reversible cycle\,:
\begin{equation}\label{Clausius2}
	\left({	\oint \frac{\diff Q}{T} }\right)_r= 0
\end{equation}
where $Q$ is the heat exchanged by the system with its environment, which sign is relative to the system (positive when received); the subscript \textquote{$r$} stands for \textquote{reversible}; and $T$ is the thermal energy (in Joule) of the system. This observation makes
\begin{equation}\label{diffS}
	\diff S = \left({\frac{\diff Q}{T}}\right)_r
\end{equation}
an exact differential that allows us to define a new state-quantity $S$ (that only depends on the state) called entropy and has no dimension (the Boltzmann constant is already incorporated in \textquote{temperature} $T$). In the sequel, $S$ can also be called Clausius-entropy to remove any ambiguity. This is the initial definition of entropy. The difference in entropy between two states A and B is given by\,:
\begin{equation}\label{Clausius3}
	S_B - S_A = \left( {\int_A^B \frac{\diff Q}{T} }\right)_r
\end{equation}
Basically, the difference in entropy between two states is equal to the net heat exchanged expressed in unit of thermal energy when the path linking them is reversible. 
This can be considered an empirical law.

\subsection{Entropy and irreversibility}\label{SecLaw2}

Reversibility is actually a limiting behavior that is never reached, but towards which quasi-static processes tend when their rate of change tends towards zero. Instead of Eq.\ref{Clausius2}, %and \ref{Clausius3}, 
for any differentiable cycle what is actually observed is that heat is always transferred to the environment 
\begin{equation}
	\oint \frac{\diff Q}{T}\le 0,
\end{equation}
so that things are never the same as before.

Instead of \ref{diffS} and \ref{Clausius3}, for any differentiable actual process, one has\,:
\begin{equation}\label{diffSbis}
	\diff S = \left({\frac{\diff Q}{T}}\right)_r \quad\ge\quad \frac{\diff Q}{T}
\end{equation}
and
\begin{equation}\label{diffS_Clausius2}
	\left[  S_{\text{B}} -S_{\text{A}}=	\displaystyle\left( {\int_A^B \frac{\diff Q}{T} }\right)_r \right]
	\quad \ge \quad 
	\left[  \displaystyle\int_A^B \frac{\diff Q}{T} = Q_{AB} \right]
\end{equation}
For non-differentiable processes, the integral on the right-hand side has to be replaced by a sum.
These are different forms of the Clausius inequality, which expresses thermodynamic irreversibility and can also be considered empirical.

With the definition of entropy (Eq.\,\ref{diffS}), Clausius inequality forms the second law of thermodynamics (the first being the principle of conservation).

\subsection{Entropy and equilibrium}\label{attractor}

Let us return to the problem of the attractiveness of equilibrium.
As is the case in classical mechanics, the problem is solved if the equilibrium is defined as the state that minimizes a potential\,\cite{Callen_1985}.
In thermodynamics, this potential always involves the negative of entropy (negentropy). 
%Hence the fact that this definition comes after that of entropy.
Its very definition depends on the interactions between the system and its environment. For example, for a thermalized system it is the free energy $F=U-TS$, where $U$ is the internal energy, etc. 

For simplicity, let us consider the union of a system of interest and its environment, the latter being chosen in such a way that the union can be considered as an isolated system. The system is in a stationary state when its interactions with the environment are also in a stationary state. So that the equilibrium of the system is also that of the union, that is to say it boils down to that of an isolated system. The definition of the equilibrium for the latter is then postulated as\,:

\law{Principle (of maximum entropy)}{The equilibrium of a thermodynamic isolated system is that of maximum entropy.}

\noindent For an isolated system, no heat is exchanged with the environment and Eq.\ref{diffS_Clausius2} turns into
\begin{equation}\label{isolated}
	S_{\text{B}} -S_{\text{A}}\ge 0
\end{equation}
So that, the entropy of an isolated system can only increase. Adding to this the principle of maximum entropy, the equilibrium becomes an attractor.

\section{The laws of thermodynamics from the fundamental principles}

As already mentioned, the second law of thermodynamics summarized in \S\ref{SecLaw1} and \ref{SecLaw2} originates from observations. It is empirical. The aim of this section is mainly to show that it can be derived from the fundamental principles listed in \S\ref{neo}. The derivation passes firstly through the contribution of statistical mechanics (Gibbs\,\cite{Gibbs_1902}), and secondly through that of information theory (Shannon\,\cite{Shannon_1948}).
The latter sheds a new light on what entropy is, making it possible to link heat (therefore energy) and information (therefore the observer).

\subsection{Entropy and improbability}

With the advent of atoms in physics comes also thermal agitation, probabilities and statistics. The system constantly changes its microscopic configuration (also named microstate, or phase). So that the equilibrium (the purpose of thermodynamics) should be understood as an average state. This is the contribution of statistical mechanics: having found a way to link microscopic to macroscopic scales by expressing state quantities, such as entropy, in terms of average quantities over microstates (for a recent textbook about the foundations of statistical mechanics see e.g. \cite{Frigg2023}). 

For the calculation, we must first decide on their probability distribution. Knowing that it cannot be measured, this distribution is  necessarily \textit{a priori} postulated. \myquote{When one does not know anything the answer is simple. One is satisfied with enumerating the possible events and assigning equal probabilities to them.} (Balian\,\cite{Balian_1991}). The distribution to be used is uniform.
This is the fundamental postulate of statistical mechanics (a variation of the Laplace's principle of insufficient reason\,\cite{Dubs_1942}).
Against this postulate Ellis wrote\,:
\myquote{It cannot be that because we are ignorant of the matter we know something about~it}\,\cite{Ellis_1850}. But according to neopositivism (\S\ref{neo}), a postulate is not knowledge so that there is actually no inconsistency.
As all conventions, this postulate has no obligation to be confirmed by direct observations. 
Only indirect feedback is required; it will be provided by the agreement between the calculated macroscopic quantities and those measured.
And it has to be said, it works\,!

Let us denote $S_\text{Gibbs}$ the average of the logarithm of improbabilities $1/p_i$, where $p_i$ is the probability of microstate $i$.
\begin{equation}\label{GS}
	S_\text{Gibbs}=\sum\limits_{i\in\Gamma} p_i\ln 1/p_i,
\end{equation}
$\Gamma$ is the ensemble of possible microstates (the phase space).
By starting with a uniform distribution of microstates for an isolated system (the whole) about which nothing is known, we can derive that of any subpart (see e.g. \cite{Lairez_2022a}), then deduce the equality\,:
\begin{equation}\label{GS2}
	S=S_\text{Gibbs}
\end{equation}
hence the name Gibbs-entropy (or statistical entropy) for $S_\text{Gibbs}$. 
This last equality, which links entropy to the improbability of microstates, is conditioned only to the postulate of their uniform distribution for the whole.

\subsection{Entropy and uncertainty}\label{uncert}

Consider each microstate identified by an integer $i\in[0,\mathcal{W}-1 ]$ (with $\mathcal W$ is the cardinality of $\Gamma$) and encode $i$ in binary. A thermodynamic system can be considered a dynamic random source of these numbers.
Independently of thermodynamics and of statistical mechanics, Shannon\,\cite{Shannon_1948} was interested in the recording space needed to store such random outputs. 
An initial idea\,\cite{Hartley_1928} was to allocate the same storage space to everyone, i.e. $\log_2\mathcal W$ bits per output in order to be able to store the largest possible integer when it comes out. But knowledge of the probability distribution of the outcomes, and the allocation of a different storage space for each, make it possible to reserve the largest identifiers for the least probable outcomes and a lossless compression of their storage.
Interestingly, Shannon showed that in no case the average number of bit per output can be less than\,:
\begin{equation}\label{ShannonH}
	H = \sum_{i} {p_i \log_2(1/p_i)} 
\end{equation}	
$H$ was named by Shannon quantity of information emitted by the source. 
Identification of Eq.\ref{ShannonH} with the Gibbs formula Eq.\ref{GS} is immediate, so that 
\begin{equation}\label{SH}
	S_\text{Shannon}=H \times \ln 2
\end{equation}
was called Shannon-entropy of the random variable under consideration. Gibbs-entropy is the Shannon-entropy of microstates.
To a factor, and only conditioned on the validity of Eq.\ref{GS2}, the thermodynamic entropy of a system turns out to be equal to the quantity of information it emits (the quantity of information of the microstates random variable).

Let us make a few remarks about the denomination \textquote{quantity of information} emitted by the system.
If we (the receiver) know nothing about the emission of the source,
it must be planed at least $\log_2 \mathcal W$ per output.
There is no possibility for lossless compression. But, if we learn that it is now certain that the first bit will not change, then it is no longer necessary to record that bit. We have received one piece of information (one information-bit) that allowed us to save one bit of storage space.
And this, for each information-bit until $H$ is zero when there is no uncertainty about the outcomes. 
Note that when a piece of information is provided to us, it  saves one bit of storage space. It is not necessary to provide this twice; the economy would not increase further. The thing called \textquote{information} that is quantified by Shannon, obeys to the principle of no redundancy and is conform to what was understood as information in \S\ref{neo}. The term \textquote{quantity of information} is thus in a certain sense appropriate. But in my opinion, it would have been clearer if the term \textquote{information} had remained reserved for a single transmitted truth, and if the term \textquote{knowledge} (instead of \textquote{quantity of information}) had been used when it came to quantifying it as the sum total of truths stored in our memory.
In any way, $H$ represents the knowledge we lack regarding the current microstate of the system. Or, from another perspective, $H$ is a measure of the uncertainty about it.

Let us go back to the problem of the \textit{a priori} probability distribution of microstates which alone conditions the validity of Eq.\ref{GS2}.
Consider the probability distribution of random outcomes of which we only have partial knowledge.
According to the principle of least-talker, the distribution must maximize the uncertainty.
Shore and Johnson\,\cite{Shore_1980} showed that maximizing $H$ (instead of another possible measure of uncertainty) is the only procedure that 
respects the principle of univocality of~\S\ref{neo}. 
Maximizing $H$ is the only procedure consistent with neopositivism to decide for any \textit{a priori} probability distribution. Hence the theorem\,:

\law{Maximum Shannon-entropy theorem (MaxEnt)}{For any random variable, the only \textit{a priori} probability distribution, compatible with the principle of least-talker and that of univocality, is the one that maximizes Shannon entropy with the constraint of taking into account our knowledge.}

\noindent In the case where we know nothing, it can be shown that the maximum of Shannon-entropy is obtained for the uniform distribution\,\cite{Shannon_1948}. Thus, the fundamental postulate of statistical mechanics does not come as an addition to the theory; it can be deduced from the fundamental principles of neopositivism and can be considered already contained in them.
And this holds \textit{de facto} for Eq.\ref{GS2} and for the equality between Clausius-entropy and Shannon-entropy of microstates. So that\,:
\begin{equation}\label{SandH}
	S=S_\text{Gibbs}=S_\text{Shannon}
\end{equation}
This equality is not conditional on any postulates other than the principle of least-talker and that of univocality.

\subsection{Clausius inequality from the principle of no redundancy}

Shannon quantity of information $H$ is actually a quantity of information we (the observer) lack about the current microstate of a thermodynamic system. Thanks to \ref{SandH}, equation \ref{SH} turns into\,:
\begin{equation}
	H = S/\ln 2
\end{equation}
For our purpose, it is clearer to speak in terms relative to us, rather than in terms relative to the system. That is to say in terms of the information we have, our knowledge $K$, rather than in terms of the information we lack, $H$. The two quantities vary in opposite ways\,:
\begin{equation}\label{K}
	\Delta K=-\Delta H = -\Delta S/\ln 2
\end{equation}
Since thermodynamics only deals with variations (differences) of entropy, the choice of the origin is not important. So that with a proper choice of origin, our knowledge $K$ can be considered, to a factor $1/\ln 2$, equal to the Brillouin's negentropy\,\cite{Brillouin_1956_book}\,: $K=-S/\ln 2$.

Consider a reversible process from $A$ to $B$. By denoting $\diff q = \diff Q/T\ln 2$, Eq.\ref{Clausius3} rewrites in terms of knowledge\,:
\begin{equation}\label{revK}
	\Delta K=	K_B-K_A = - \left( {\int_A^B \diff q }\right)_r	
\end{equation}
For simplicity, we will consider a discrete case allowing to rewrite the above equation\,:
\begin{equation}\label{revK2}
	\Delta K  = \left(\mathcal Q\right)_r \quad
	\text{with} \quad \mathcal Q = -\sum_A^B q_i
\end{equation}
The above equality holds for the reversible case, our purpose is to examine the order relationship between these two terms in the general case.

\begin{itemize}[]
	\item The term on the right refers to the system only. $\mathcal Q$ is the total amount of heat (in unit of $T\ln 2$, i.e. it has no dimension) emitted during the process by the system into the entire environment, of which the observer is a part.
	
	\item The term on the left refers to the observer only. It represents the variation in our knowledge about microstates.
\end{itemize}

\noindent Consider the process the system undergoes is accompanied by the sequential emission of discrete quantities of heat $q_i$ at regular time interval~$\tau$ (which can be viewed as the characteristic time of heat diffusion through the system).
The system is an emitter of data under the form of heat (not to be confused with the emission of a random variable evoked in \S\ref{uncert}). While the observer can be considered as a receiver sequentially sampling the emission. This sampling is also characterized by a time that we denote $\tau_s$. The quantity of data sampled and received by the observer is\,:
\begin{equation}
	D = -\sum_A^B q_j
\end{equation}
Let us examine the three possibilities\,: 

\begin{enumerate}
	\item $\tau_s = \tau$\,:
	
	The number of data emitted and received is identical, so that $D=\mathcal Q$. Each data $q_j$ is a piece of information and increases knowledge by the same quantity. So that\,:
	
	\begin{equation}
		\Delta K_{(\tau_s = \tau)} = D = \mathcal Q
	\end{equation}

	\item $\tau_s < \tau$ (oversampling)
	
	The observer acquires redundant data. According to the principle of no redundancy, these do not increase knowledge and count as one (the oversampling is useless in terms of information). So that\,:
	
	\begin{equation}
		\Delta K_{(\tau_s < \tau)} = D = \mathcal Q
	\end{equation}
	
	\item $\tau_s > \tau$ (undersampling)
	
	According to the first principle of truth and from the definition of knowledge, $K$~can only increase with observed data. But in the case of undersampling, the observer is missing data $D<\mathcal Q$. So that:
	$$
	\Delta K_{(\tau_s > \tau)} = D < \mathcal Q
	$$
\end{enumerate}
The first two cases correspond to reversible processes, the third to irreversible ones. In any cases we have the inequality\,:

\begin{equation}
	\Delta K \quad \le \quad \mathcal Q
\end{equation}
which is nothing but the Clausius inequality. This stems from the postulates that information is a truth (definition) that necessarily requires observation (principle of truth 1/2), but cannot be given twice (principle of no redundancy).

The result can be reformulated in everyday language that makes it quite obvious. Increasing knowledge requires new observations, but in no case can the amount of observed things be greater than the amount of observable things. It is also worth noting that a parallel between Clausius inequality and the Nyquist–Shannon sampling theorem in signal processing is quite relevant\,: oversampling does not increase information, but undersampling acts as a low-pass filter. If these reformulations of Clausius inequality seem obvious to us, it is actually because we are already steeped in neopositivist thinking.

\subsection{Knowledge and equilibrium}

The last point to examine is that of the definition of thermodynamic equilibrium. The need for it to be an attractor led us in \S\ref{attractor} to define it as the state of maximum entropy. But this can be considered as an additional postulate.
Maximum Shannon-entropy theorem (\S\ref{uncert}) potentially gives us the possibility of deducing this definition, thus making the economy of a postulate. 

However, the Maximum Shannon-entropy theorem does not tell us the distribution of which random variable should be maximized at the thermodynamic equilibrium. Different random variables may lead to different incompatible definitions of the equilibrium, inconsistently with the principle of univocality.
Jaynes\,\cite{Jaynes_1973} note the following point. Consider the three physical symmetries\,: translation, rotation and scaling. Any definition of the equilibrium that is not invariant under these symmetries would actually be equivocal. Or in other words, any definition of equilibrium that depends on these symmetries would say more about equilibrium than they should, inconsistently with the least-talker principle.
Actually, a definition of thermodynamic equilibrium that is consistent with the principle of least-talker and that of univocality must maximize Shannon-entropy of random variables which distribution is invariant in form upon these symmetries. Examples of these variables include local density and microstate. In virtue of Eq.\ref{SandH} (that also only depends on the same principles), maximizing Shannon-entropy of these variables is thus equivalent to maximizing Clausius-entropy of the system.

The definition of thermodynamic equilibrium is also already contained in the fundamental principles of neopositivism. It can be derived from the fundamental principle of least-talker and that of univocality.

\section{Conclusion}

In thermodynamics a central state quantity of a system, namely its entropy, is related to our knowledge. A state quantity of a system is in fact also a state quantity of the observer.
This was already recognized long before Shannon. In 1878, Maxwell wrote \myquote{The idea of dissipation of energy depends on the extent of our knowledge}\,\cite{Maxwell_1878}.
It is therefore not surprising and likely inevitable to relate thermodynamics to epistemology, i.e. the branch of philosophy which deals with knowledge. 
Fundamental principles of neopositivism concerning what reality is, what the sole source of truth is, what knowledge is, in fact allow us to deduce the second law of thermodynamics, that of the irreversibility of phenomena, this law which makes thermodynamics so special compared to all other theories of physics.
Fundamental postulates cannot be \textit{a priori} justified. Their justification lies solely in what they allow us to deduce by comparison with observations, by comparison with empirical laws. In our case, from neopositivism, we can deduce the second law of thermodynamics (an empirical law). Here is its justification.
Interestingly, these fundamental principles are not special to thermodynamics, but apply to all fields of science and in particular to that of physics. In my opinion, unification of physics lies on this side.

\myquote{A reality completely independent of the mind which conceives it, sees or feels it, is an impossibility. A world as exterior as that, even if it existed,	would for us be forever inaccessible. But what we call objective reality is, in the last analysis, what is common to many thinking beings, and could be common to all} (Poincar\'e\,\cite{Poincare_1907}).

\bibliography{\string~/Documents/Articles/weri_biblio.bib}

\begin{thebibliography}{31}%
\makeatletter
\providecommand \@ifxundefined [1]{%
 \@ifx{#1\undefined}
}%
\providecommand \@ifnum [1]{%
 \ifnum #1\expandafter \@firstoftwo
 \else \expandafter \@secondoftwo
 \fi
}%
\providecommand \@ifx [1]{%
 \ifx #1\expandafter \@firstoftwo
 \else \expandafter \@secondoftwo
 \fi
}%
\providecommand \natexlab [1]{#1}%
\providecommand \enquote  [1]{``#1''}%
\providecommand \bibnamefont  [1]{#1}%
\providecommand \bibfnamefont [1]{#1}%
\providecommand \citenamefont [1]{#1}%
\providecommand \href@noop [0]{\@secondoftwo}%
\providecommand \href [0]{\begingroup \@sanitize@url \@href}%
\providecommand \@href[1]{\@@startlink{#1}\@@href}%
\providecommand \@@href[1]{\endgroup#1\@@endlink}%
\providecommand \@sanitize@url [0]{\catcode `\\12\catcode `\$12\catcode
  `\&12\catcode `\#12\catcode `\^12\catcode `\_12\catcode `\%12\relax}%
\providecommand \@@startlink[1]{}%
\providecommand \@@endlink[0]{}%
\providecommand \url  [0]{\begingroup\@sanitize@url \@url }%
\providecommand \@url [1]{\endgroup\@href {#1}{\urlprefix }}%
\providecommand \urlprefix  [0]{URL }%
\providecommand \Eprint [0]{\href }%
\providecommand \doibase [0]{https://doi.org/}%
\providecommand \selectlanguage [0]{\@gobble}%
\providecommand \bibinfo  [0]{\@secondoftwo}%
\providecommand \bibfield  [0]{\@secondoftwo}%
\providecommand \translation [1]{[#1]}%
\providecommand \BibitemOpen [0]{}%
\providecommand \bibitemStop [0]{}%
\providecommand \bibitemNoStop [0]{.\EOS\space}%
\providecommand \EOS [0]{\spacefactor3000\relax}%
\providecommand \BibitemShut  [1]{\csname bibitem#1\endcsname}%
\let\auto@bib@innerbib\@empty
%</preamble>
\bibitem [{\citenamefont {Feynman}(1963)}]{Feynman_1963_symmetry}%
  \BibitemOpen
  \bibfield  {author} {\bibinfo {author} {\bibfnamefont {R.}~\bibnamefont
  {Feynman}},\ }\href {https://www.feynmanlectures.caltech.edu/I_52.html}
  {\bibinfo {title} {The {F}eynman lectures on physics vol. {I}, chap. 52:
  Symmetry in physical laws}} (\bibinfo {year} {1963})\BibitemShut {NoStop}%
\bibitem [{\citenamefont {Poincar\'e}(1905)}]{Poincare_1905_Sci_Hypo}%
  \BibitemOpen
  \bibfield  {author} {\bibinfo {author} {\bibfnamefont {H.}~\bibnamefont
  {Poincar\'e}},\ }\href {https://www.gutenberg.org/ebooks/37157} {\emph
  {\bibinfo {title} {Science and hypothesis}}}\ (\bibinfo  {publisher} {The
  Walter Scott Publishing Co.},\ \bibinfo {year} {1905})\BibitemShut {NoStop}%
\bibitem [{\citenamefont {Carnap}(1966)}]{Carnap_1966}%
  \BibitemOpen
  \bibfield  {author} {\bibinfo {author} {\bibfnamefont {R.}~\bibnamefont
  {Carnap}},\ }\href@noop {} {\emph {\bibinfo {title} {Philosophical
  foundations of physics}}}\ (\bibinfo  {publisher} {Basic books, New York},\
  \bibinfo {year} {1966})\BibitemShut {NoStop}%
\bibitem [{\citenamefont {Poincar\'e}(1907)}]{Poincare_1907}%
  \BibitemOpen
  \bibfield  {author} {\bibinfo {author} {\bibfnamefont {H.}~\bibnamefont
  {Poincar\'e}},\ }\href {https://www3.nd.edu/~powers/ame.60611/poincare.pdf}
  {\emph {\bibinfo {title} {The value of science}}}\ (\bibinfo  {publisher}
  {The science Press},\ \bibinfo {year} {1907})\BibitemShut {NoStop}%
\bibitem [{\citenamefont {Hahn}\ \emph {et~al.}(1973)\citenamefont {Hahn},
  \citenamefont {Neurath},\ and\ \citenamefont {Carnap}}]{Neurath1973}%
  \BibitemOpen
  \bibfield  {author} {\bibinfo {author} {\bibfnamefont {H.}~\bibnamefont
  {Hahn}}, \bibinfo {author} {\bibfnamefont {O.}~\bibnamefont {Neurath}},\ and\
  \bibinfo {author} {\bibfnamefont {R.}~\bibnamefont {Carnap}},\ }\bibinfo
  {title} {The scientific conception of the world: The {V}ienna circle},\ in\
  \href {https://doi.org/10.1007/978-94-010-2525-6_9} {\emph {\bibinfo
  {booktitle} {Empiricism and Sociology}}}\ (\bibinfo  {publisher} {Springer
  Netherlands},\ \bibinfo {address} {Dordrecht},\ \bibinfo {year} {1973})\ pp.\
  \bibinfo {pages} {299--318}\BibitemShut {NoStop}%
\bibitem [{\citenamefont {Wittgenstein}(1922)}]{Wittgenstein1922}%
  \BibitemOpen
  \bibfield  {author} {\bibinfo {author} {\bibfnamefont {L.}~\bibnamefont
  {Wittgenstein}},\ }\bibinfo {title} {Tractatus logico-philosophicus}\
  (\bibinfo  {publisher} {Kegan Paul, Trench, Trubner and Co. Ltd.},\ \bibinfo
  {year} {1922})\BibitemShut {NoStop}%
\bibitem [{\citenamefont {Mach}(1911)}]{Mach_1911}%
  \BibitemOpen
  \bibfield  {author} {\bibinfo {author} {\bibfnamefont {E.}~\bibnamefont
  {Mach}},\ }\href {https://doi.org/10.1017/cbo9781107338746} {\emph {\bibinfo
  {title} {History and root of the principle of the conservation of energy}}}\
  (\bibinfo  {publisher} {The open court publishing},\ \bibinfo {year}
  {1911})\BibitemShut {NoStop}%
\bibitem [{\citenamefont {Mach}(1919)}]{Mach_1919}%
  \BibitemOpen
  \bibfield  {author} {\bibinfo {author} {\bibfnamefont {E.}~\bibnamefont
  {Mach}},\ }\href@noop {} {\emph {\bibinfo {title} {The Science of
  Mechanics}}}\ (\bibinfo  {publisher} {The Open Court Publishing Company},\
  \bibinfo {year} {1919})\BibitemShut {NoStop}%
\bibitem [{\citenamefont {Einstein}(1934)}]{Einstein_1934}%
  \BibitemOpen
  \bibfield  {author} {\bibinfo {author} {\bibfnamefont {A.}~\bibnamefont
  {Einstein}},\ }\bibfield  {title} {\bibinfo {title} {On the method of
  theoretical physics},\ }\href {https://doi.org/10.1086/286316} {\bibfield
  {journal} {\bibinfo  {journal} {Philosophy of Science}\ }\textbf {\bibinfo
  {volume} {1}},\ \bibinfo {pages} {163} (\bibinfo {year} {1934})}\BibitemShut
  {NoStop}%
\bibitem [{\citenamefont {Ichikawa}\ and\ \citenamefont
  {Steup}(2026)}]{Ichikawa2026}%
  \BibitemOpen
  \bibfield  {author} {\bibinfo {author} {\bibfnamefont {J.}~\bibnamefont
  {Ichikawa}}\ and\ \bibinfo {author} {\bibfnamefont {M.}~\bibnamefont
  {Steup}},\ }\bibfield  {title} {\bibinfo {title} {{The Analysis of
  Knowledge}},\ }in\ \href
  {https://plato.stanford.edu/archives/sum2026/entries/knowledge-analysis/}
  {\emph {\bibinfo {booktitle} {The {Stanford} Encyclopedia of Philosophy}}},\
  \bibinfo {editor} {edited by\ \bibinfo {editor} {\bibfnamefont {E.~N.}\
  \bibnamefont {Zalta}}\ and\ \bibinfo {editor} {\bibfnamefont
  {U.}~\bibnamefont {Nodelman}}}\ (\bibinfo  {publisher} {Metaphysics Research
  Lab, Stanford University},\ \bibinfo {year} {2026})\ \bibinfo {edition}
  {{S}ummer 2026}\ ed.\BibitemShut {Stop}%
\bibitem [{\citenamefont {Lutz}\ and\ \citenamefont
  {Ciliberto}(2015)}]{Lutz_2015}%
  \BibitemOpen
  \bibfield  {author} {\bibinfo {author} {\bibfnamefont {E.}~\bibnamefont
  {Lutz}}\ and\ \bibinfo {author} {\bibfnamefont {S.}~\bibnamefont
  {Ciliberto}},\ }\bibfield  {title} {\bibinfo {title} {Information: from
  {M}axwell's demon to {L}andauer's eraser},\ }\href
  {https://doi.org/10.1063/pt.3.2912} {\bibfield  {journal} {\bibinfo
  {journal} {Physics Today}\ }\textbf {\bibinfo {volume} {68}},\ \bibinfo
  {pages} {30} (\bibinfo {year} {2015})}\BibitemShut {NoStop}%
\bibitem [{\citenamefont {Ciliberto}\ and\ \citenamefont
  {Lutz}(2018)}]{Ciliberto_2018}%
  \BibitemOpen
  \bibfield  {author} {\bibinfo {author} {\bibfnamefont {S.}~\bibnamefont
  {Ciliberto}}\ and\ \bibinfo {author} {\bibfnamefont {E.}~\bibnamefont
  {Lutz}},\ }\bibfield  {title} {\bibinfo {title} {The physics of information:
  from {M}axwell to {L}andauer},\ }in\ \href
  {https://doi.org/10.1007/978-3-319-93458-7_5} {\emph {\bibinfo {booktitle}
  {Energy Limits in Computation}}}\ (\bibinfo  {publisher} {Springer
  International Publishing},\ \bibinfo {year} {2018})\ pp.\ \bibinfo {pages}
  {155--175}\BibitemShut {NoStop}%
\bibitem [{\citenamefont {Landauer}(1961)}]{Landauer_1961}%
  \BibitemOpen
  \bibfield  {author} {\bibinfo {author} {\bibfnamefont {R.}~\bibnamefont
  {Landauer}},\ }\bibfield  {title} {\bibinfo {title} {Irreversibility and heat
  generation in the computing process},\ }\href
  {https://doi.org/10.1147/rd.53.0183} {\bibfield  {journal} {\bibinfo
  {journal} {{IBM} Journal of Research and Development}\ }\textbf {\bibinfo
  {volume} {5}},\ \bibinfo {pages} {183} (\bibinfo {year} {1961})}\BibitemShut
  {NoStop}%
\bibitem [{\citenamefont {Lairez}(2025)}]{Lairez2025}%
  \BibitemOpen
  \bibfield  {author} {\bibinfo {author} {\bibfnamefont {D.}~\bibnamefont
  {Lairez}},\ }\bibfield  {title} {\bibinfo {title} {Disentangling
  {B}rillouin’s negentropy law of information and {L}andauer’s law on data
  erasure},\ }\href {https://doi.org/10.3390/e28010037} {\bibfield  {journal}
  {\bibinfo  {journal} {Entropy}\ }\textbf {\bibinfo {volume} {28}},\ \bibinfo
  {pages} {37} (\bibinfo {year} {2025})}\BibitemShut {NoStop}%
\bibitem [{\citenamefont {Hahn}(1980)}]{Hahn1980}%
  \BibitemOpen
  \bibfield  {author} {\bibinfo {author} {\bibfnamefont {H.}~\bibnamefont
  {Hahn}},\ }\bibinfo {title} {Superfluous entities, or {O}ccam’s razor},\
  in\ \href {https://doi.org/10.1007/978-94-009-8982-5_1} {\emph {\bibinfo
  {booktitle} {Empiricism, Logic and Mathematics}}}\ (\bibinfo  {publisher}
  {Springer Netherlands},\ \bibinfo {year} {1980})\ pp.\ \bibinfo {pages}
  {1--19}\BibitemShut {NoStop}%
\bibitem [{\citenamefont {Clausius}(1879)}]{Clausius_1879}%
  \BibitemOpen
  \bibfield  {author} {\bibinfo {author} {\bibfnamefont {R.}~\bibnamefont
  {Clausius}},\ }\href
  {https://books.google.fr/books?id=8LIEAAAAYAAJ&printsec=frontcover&hl=fr&source=gbs_ge_summary_r&cad=0#v=onepage&q&f=false}
  {\emph {\bibinfo {title} {The mechanical theory of heat}}}\ (\bibinfo
  {publisher} {Macmillan \& Co, London, UK},\ \bibinfo {year}
  {1879})\BibitemShut {NoStop}%
\bibitem [{\citenamefont {Feynman}\ \emph {et~al.}(1966)\citenamefont
  {Feynman}, \citenamefont {Leighton},\ and\ \citenamefont
  {Sands}}]{Feynmann_Energy}%
  \BibitemOpen
  \bibfield  {author} {\bibinfo {author} {\bibfnamefont {R.~P.}\ \bibnamefont
  {Feynman}}, \bibinfo {author} {\bibfnamefont {R.~B.}\ \bibnamefont
  {Leighton}},\ and\ \bibinfo {author} {\bibfnamefont {M.}~\bibnamefont
  {Sands}},\ }\href {https://www.feynmanlectures.caltech.edu/I_04.html} {\emph
  {\bibinfo {title} {The {F}eynman lectures on physics}}}\ (\bibinfo
  {publisher} {Addison-Wesley, Reading, MA},\ \bibinfo {year} {1966})\
  Chap.~\bibinfo {chapter} {4}\BibitemShut {NoStop}%
\bibitem [{\citenamefont {Callen}(1985)}]{Callen_1985}%
  \BibitemOpen
  \bibfield  {author} {\bibinfo {author} {\bibfnamefont {H.~B.}\ \bibnamefont
  {Callen}},\ }\href@noop {} {\emph {\bibinfo {title} {Thermodynamics and an
  introduction to thermostatistics}}},\ \bibinfo {edition} {2nd}\ ed.\
  (\bibinfo  {publisher} {J. Wiley \& sons},\ \bibinfo {year}
  {1985})\BibitemShut {NoStop}%
\bibitem [{\citenamefont {Clausius}(1865)}]{Clausius_1865}%
  \BibitemOpen
  \bibfield  {author} {\bibinfo {author} {\bibfnamefont {R.}~\bibnamefont
  {Clausius}},\ }\bibfield  {title} {\bibinfo {title} {Sur diverses formes
  facilement applicables qu'on peut donner aux {\'e}quations fondamentales de
  la th{\'e}orie m{\'e}canique de la chaleur.},\ }\href
  {http://eudml.org/doc/235530} {\bibfield  {journal} {\bibinfo  {journal}
  {Journal de Math{\'e}matiques Pures et Appliqu{\'e}es}\ ,\ \bibinfo {pages}
  {361}} (\bibinfo {year} {1865})}\BibitemShut {NoStop}%
\bibitem [{\citenamefont {Gibbs}(1902)}]{Gibbs_1902}%
  \BibitemOpen
  \bibfield  {author} {\bibinfo {author} {\bibfnamefont {J.}~\bibnamefont
  {Gibbs}},\ }\href {https://www.gutenberg.org/files/50992/50992-pdf.pdf}
  {\emph {\bibinfo {title} {Elementary principles in statistical mechanics}}}\
  (\bibinfo  {publisher} {Charles Scribner's sons},\ \bibinfo {year}
  {1902})\BibitemShut {NoStop}%
\bibitem [{\citenamefont {Shannon}(1948)}]{Shannon_1948}%
  \BibitemOpen
  \bibfield  {author} {\bibinfo {author} {\bibfnamefont {C.~E.}\ \bibnamefont
  {Shannon}},\ }\bibfield  {title} {\bibinfo {title} {A mathematical theory of
  communication},\ }\href {https://doi.org/10.1002/j.1538-7305.1948.tb01338.x}
  {\bibfield  {journal} {\bibinfo  {journal} {The Bell System Technical
  Journal}\ }\textbf {\bibinfo {volume} {27}},\ \bibinfo {pages} {379}
  (\bibinfo {year} {1948})}\BibitemShut {NoStop}%
\bibitem [{\citenamefont {Frigg}\ and\ \citenamefont
  {Werndl}(2023)}]{Frigg2023}%
  \BibitemOpen
  \bibfield  {author} {\bibinfo {author} {\bibfnamefont {R.}~\bibnamefont
  {Frigg}}\ and\ \bibinfo {author} {\bibfnamefont {C.}~\bibnamefont {Werndl}},\
  }\href {https://doi.org/10.1017/9781009022798} {\emph {\bibinfo {title}
  {Foundations of Statistical Mechanics}}}\ (\bibinfo  {publisher} {Cambridge
  University Press},\ \bibinfo {year} {2023})\BibitemShut {NoStop}%
\bibitem [{\citenamefont {Balian}(1991)}]{Balian_1991}%
  \BibitemOpen
  \bibfield  {author} {\bibinfo {author} {\bibfnamefont {R.}~\bibnamefont
  {Balian}},\ }\href {https://doi.org/10.1007/978-3-540-45475-5} {\emph
  {\bibinfo {title} {From microphysics to macrophysics}}}\ (\bibinfo
  {publisher} {Springer Berlin Heidelberg},\ \bibinfo {year}
  {1991})\BibitemShut {NoStop}%
\bibitem [{\citenamefont {Dubs}(1942)}]{Dubs_1942}%
  \BibitemOpen
  \bibfield  {author} {\bibinfo {author} {\bibfnamefont {H.}~\bibnamefont
  {Dubs}},\ }\bibfield  {title} {\bibinfo {title} {The principle of
  insufficient reason},\ }\href {https://doi.org/10.1086/286754} {\bibfield
  {journal} {\bibinfo  {journal} {Philosophy of Science}\ }\textbf {\bibinfo
  {volume} {9}},\ \bibinfo {pages} {123} (\bibinfo {year} {1942})}\BibitemShut
  {NoStop}%
\bibitem [{\citenamefont {Ellis}(1850)}]{Ellis_1850}%
  \BibitemOpen
  \bibfield  {author} {\bibinfo {author} {\bibfnamefont {R.}~\bibnamefont
  {Ellis}},\ }\bibfield  {title} {\bibinfo {title} {Remarks on an alleged proof
  of the {\textquotedblleft}method of least squares,{\textquotedblright}
  contained in a late number of the edinburgh review},\ }\href
  {https://doi.org/10.1080/14786445008646622} {\bibfield  {journal} {\bibinfo
  {journal} {The London, Edinburgh, and Dublin Philosophical Magazine and
  Journal of Science}\ }\textbf {\bibinfo {volume} {37}},\ \bibinfo {pages}
  {321} (\bibinfo {year} {1850})}\BibitemShut {NoStop}%
\bibitem [{\citenamefont {Lairez}(2022)}]{Lairez_2022a}%
  \BibitemOpen
  \bibfield  {author} {\bibinfo {author} {\bibfnamefont {D.}~\bibnamefont
  {Lairez}},\ }\bibfield  {title} {\bibinfo {title} {A short derivation of
  {B}oltzmann distribution and {G}ibbs entropy formula from the fundamental
  postulate},\ }\href@noop {} {\bibfield  {journal} {\bibinfo  {journal}
  {arXiv}\ } (\bibinfo {year} {2022})},\ \Eprint
  {https://arxiv.org/abs/2211.02455v3} {2211.02455v3} \BibitemShut {NoStop}%
\bibitem [{\citenamefont {Hartley}(1928)}]{Hartley_1928}%
  \BibitemOpen
  \bibfield  {author} {\bibinfo {author} {\bibfnamefont {R.~V.~L.}\
  \bibnamefont {Hartley}},\ }\bibfield  {title} {\bibinfo {title} {Transmission
  of information},\ }\href {https://doi.org/10.1002/j.1538-7305.1928.tb01236.x}
  {\bibfield  {journal} {\bibinfo  {journal} {Bell System Technical Journal}\
  }\textbf {\bibinfo {volume} {7}},\ \bibinfo {pages} {535} (\bibinfo {year}
  {1928})}\BibitemShut {NoStop}%
\bibitem [{\citenamefont {Shore}\ and\ \citenamefont
  {Johnson}(1980)}]{Shore_1980}%
  \BibitemOpen
  \bibfield  {author} {\bibinfo {author} {\bibfnamefont {J.}~\bibnamefont
  {Shore}}\ and\ \bibinfo {author} {\bibfnamefont {R.}~\bibnamefont
  {Johnson}},\ }\bibfield  {title} {\bibinfo {title} {Axiomatic derivation of
  the principle of maximum entropy and the principle of minimum
  cross-entropy},\ }\href {https://doi.org/10.1109/tit.1980.1056144} {\bibfield
   {journal} {\bibinfo  {journal} {{IEEE} Transactions on Information Theory}\
  }\textbf {\bibinfo {volume} {26}},\ \bibinfo {pages} {26} (\bibinfo {year}
  {1980})}\BibitemShut {NoStop}%
\bibitem [{\citenamefont {Brillouin}(1956)}]{Brillouin_1956_book}%
  \BibitemOpen
  \bibfield  {author} {\bibinfo {author} {\bibfnamefont {L.}~\bibnamefont
  {Brillouin}},\ }\href@noop {} {\emph {\bibinfo {title} {Science and
  Information Theory}}}\ (\bibinfo  {publisher} {Dover Publications},\ \bibinfo
  {address} {Mineola, N.Y.},\ \bibinfo {year} {1956})\BibitemShut {NoStop}%
\bibitem [{\citenamefont {Jaynes}(1973)}]{Jaynes_1973}%
  \BibitemOpen
  \bibfield  {author} {\bibinfo {author} {\bibfnamefont {E.~T.}\ \bibnamefont
  {Jaynes}},\ }\bibfield  {title} {\bibinfo {title} {The well-posed problem},\
  }\href {https://doi.org/10.1007/bf00709116} {\bibfield  {journal} {\bibinfo
  {journal} {Foundations of Physics}\ }\textbf {\bibinfo {volume} {3}},\
  \bibinfo {pages} {477} (\bibinfo {year} {1973})}\BibitemShut {NoStop}%
\bibitem [{\citenamefont {Maxwell}(1878)}]{Maxwell_1878}%
  \BibitemOpen
  \bibfield  {author} {\bibinfo {author} {\bibfnamefont {J.~C.}\ \bibnamefont
  {Maxwell}},\ }\bibfield  {title} {\bibinfo {title} {Diffusion},\ }\href
  {https://doi.org/10.1017/CBO9780511710377.064} {\bibfield  {journal}
  {\bibinfo  {journal} {Encyclopedia Britannica, reproduced in Scientific
  papers}\ }\textbf {\bibinfo {volume} {2}},\ \bibinfo {pages} {625} (\bibinfo
  {year} {1878})}\BibitemShut {NoStop}%
\end{thebibliography}%

\end{document}